\documentclass[a4paper]{article}

\usepackage{INTERSPEECH2020}
\usepackage{math}
\usepackage{amsmath}
\usepackage{listings}  
\usepackage{footnote}  
\usepackage{hyperref}  
\usepackage[super]{nth}  
\usepackage{bbding}  
\usepackage{url}  
\usepackage{cite}

\usepackage{xcolor}

\definecolor{codegreen}{rgb}{0,0.6,0}
\definecolor{codegray}{rgb}{0.5,0.5,0.5}
\definecolor{codepurple}{rgb}{0.58,0,0.82}
\definecolor{backcolour}{rgb}{0.95,0.95,0.92}

\lstdefinestyle{tree}{
    literate=
    {├}{{\smash{\raisebox{-1ex}{\rule{1pt}{\baselineskip}}}\raisebox{0.5ex}{\rule{1ex}{1pt}}}}1 
    {│}{{\smash{\raisebox{-1ex}{\rule{1pt}{\baselineskip}}}\raisebox{0.5ex}{\rule{1ex}{0pt}}}}1
    {─}{{\raisebox{0.5ex}{\rule{1.5ex}{1pt}}}}1 
    {└}{{\smash{\raisebox{0.5ex}{\rule{1pt}{\dimexpr\baselineskip-1.5ex}}}\raisebox{0.5ex}{\rule{1ex}{1pt}}}}1 
  }
  
\lstdefinestyle{mystyle}{
    backgroundcolor=\color{backcolour},   
    commentstyle=\color{codegreen},
    keywordstyle=\color{magenta},
    numberstyle=\tiny\color{codegray},
    stringstyle=\color{codepurple},
    basicstyle=\ttfamily\footnotesize,
    breakatwhitespace=false,         
    breaklines=true,                 
    captionpos=b,                    
    keepspaces=true,                 
    showspaces=false,                
    showstringspaces=false,
    showtabs=false,                  
    tabsize=2
}

\usepackage{graphicx}  
\graphicspath{{./img/}}
\usepackage{float}

\newcommand{\code}[1]{\texttt{\footnotesize{#1}}}
\newcommand{\Test}[1]{\expandafter\hat#1}

\newcommand*\Asteroid{\texttt{\footnotesize{Asteroid}}}

\title{Asteroid: the PyTorch-based audio source separation toolkit for researchers}
\name{Manuel Pariente$^1$, Samuele Cornell$^2$, Joris Cosentino$^1$, Sunit Sivasankaran$^1$, Efthymios Tzinis$^3$,\\
Jens Heitkaemper$^4$, Michel Olvera$^1$, Fabian-Robert St\"{o}ter$^5$, Mathieu Hu$^1$, Juan M. Mart\'in-Do\~{n}as$^6$,
David Ditter$^7$, 
Ariel Frank$^8$, Antoine Deleforge$^1$, Emmanuel Vincent$^1$
\thanks{We would like to thank Herv\'e Bredin for fruitful discussions about software design and Kaituo Xu 
for open-sourcing his Conv-Tasnet implementation.}
\thanks{Experiments presented in this paper were partially carried out using the Grid'5000 testbed, supported
by a scientific interest group hosted by Inria and including CNRS, RENATER and several Universities as well
as other organizations (see \texttt{https://www.grid5000.fr}).}
\thanks{High Performance Computing resources were partially provided by the EXPLOR centre hosted by the 
University de Lorraine}}
\address{$^1$Universit\'e de Lorraine, CNRS, Inria, LORIA, France $^2$Universit\`{a} Politecnica delle Marche, Italy \\
$^3$University of Illinois at Urbana-Champaign, USA
$^4$Universit\"at Paderborn, Germany\\
$^5$Inria and LIRMM, University of Montpellier, France
$^6$Universidad de Granada, Spain\\
$^7$Universit\"at Hamburg, Germany
$^8$Technion - Israel Institute of Technology, Israel}
\email{\large\href{https://github.com/mpariente/asteroid}{https://github.com/mpariente/asteroid}}

\begin{document}

    \maketitle
    \begin{abstract}
        This paper describes \Asteroid, the \code{PyTorch}-based audio source separation toolkit for researchers. 
        Inspired by the most successful neural source separation systems, it provides all neural building blocks
        required to build such a system. To improve reproducibility, Kaldi-style recipes on common audio source separation 
        datasets are also provided. This paper describes the software architecture of \code{Asteroid} and
        its most important features. By showing experimental results obtained with \code{Asteroid}'s recipes, we
        show that our implementations are at least on par with most results reported in reference papers. The toolkit
        is publicly available at \href{https://github.com/mpariente/asteroid}{github.com/mpariente/asteroid}.
    \end{abstract}

    \section{Introduction}\label{sec:introduction}

    Audio source separation, which aims to separate a mixture signal into individual source signals, is essential 
    to robust speech processing in real-world acoustic environments \cite{BookEVincent}. 
    Classical open-source toolkits such as FASST \cite{FASST2014}, HARK \cite{HARK2009},
    ManyEars \cite{ManyEars2013} and openBliSSART \cite{OpenBLISSART} which are based on probabilistic 
    modelling, non-negative matrix factorization
    , sound source localization and/or beamforming
    have been successful in the past decade. However, they are now largely outperformed by deep learning-based approaches, at least on the task of single-channel source separation \cite{DPCLHershey2016, PITYu2016, LSTMLuo2018, ConvLuo2018,
    Wavesplit2020Zeghidour}.

    Several open-source toolkits have emerged for deep learning-based source separation. These include 
    \code{nussl} (Northwestern University Source Separation Library) \cite{NUSSLManilow2018},
    \code{ONSSEN} (An Open-source Speech Separation and Enhancement Library) \cite{OnssenNi2019}, 
    \code{Open-Unmix} \cite{OpenUnmix}, and countless isolated implementations replicating some important papers 
    \footnote{\href{https://github.com/kaituoxu/TasNet}{kaituoxu/TasNet}, \href{https://github.com/kaituoxu/Conv-TasNet}{kaituoxu/Conv-TasNet},
    \href{https://github.com/yluo42/TAC}{yluo42/TAC}, \href{https://github.com/JusperLee/Conv-TasNet}{JusperLee/Conv-TasNet},
    \href{https://github.com/JusperLee/Dual-Path-RNN-Pytorch}{JusperLee/Dual-Path-RNN-Pytorch}, 
    \href{https://github.com/tky1117/DNN-based_source_separation}{tky1117/DNN-based\_source\_separation}
    \href{https://github.com/ShiZiqiang/dual-path-RNNs-DPRNNs-based-speech-separation}{ShiZiqiang/dual-path-RNNs-DPRNNs-based-speech-separation}}.

    Both \code{nussl} and \code{ONSSEN} are written in \code{PyTorch}\cite{PyTorchPaszke2019} and provide training and evaluation scripts
    for several state-of-the art methods. However, data preparation steps are not provided and experiments are not easily 
    configurable from the command line. \code{Open-Unmix} does provides a complete pipeline from data preparation until evaluation, but 
    only for the Open-Unmix model on the music source separation task.
    Regarding the isolated implementations, some of them only contain the 
    model, while others provide training scripts but assume that training data has been generated. Finally, 
    very few provide the complete pipeline. Among the ones providing 
    evaluation scripts, differences can often be found, e.g., discarding short utterances
    or splitting utterances in chunks and discarding the last one. 

    This paper describes \code{Asteroid} (Audio source separation on Steroids), a new open-source toolkit for
    deep learning-based audio source separation and speech enhancement, designed for researchers and practitioners.
    Based on \code{PyTorch}, one of the most widely used dynamic neural network toolkits, \code{Asteroid}
    is meant to be user-friendly, easily extensible, to promote reproducible research, and to enable easy experimentation. 
    As such, it supports a wide range of datasets and architectures, and comes with recipes reproducing some important papers.
    \code{Asteroid} is built on the following principles:
    \begin{enumerate} \label{guideline}
        \item Abstract only where necessary, i.e., use as much native \code{PyTorch} code as possible.
        \item Allow importing third-party code with minimal changes.
        \item Provide all steps from data preparation to evaluation.
        \item Enable recipes to be configurable from the command line.
    \end{enumerate}
    We present the audio source separation framework in Section \ref{sec:general-framework}. We describe 
    \code{Asteroid}'s main features in Section \ref{sec:functionality} and their implementation in Section \ref{sec:impl}. We provide example experimental results in Section \ref{sec:experiments} and conclude in Section \ref{sec:conclusion}.

    \section{General framework}\label{sec:general-framework}
    While \code{Asteroid} is not limited to a single task, single-channel source separation is currently its main focus.
    Hence, we will only consider this task in the rest of the paper.
     Let $x$ be a single channel recording of $J$ sources in noise:
    \begin{equation}
        \label{eq:mixing}
        x(t) = \sum_{j=1}^{J} s_{j}(t) + n(t), 
    \end{equation}
    where $\{s_j\}_{j=1..J}$ are the source signals and $n$ is an additive noise signal. The goal of source separation is to 
    obtain source estimates $\{\widehat{s}_j\}_{j=1..J}$ given $x$.

\begin{figure}[h!]
        \centering
        \includegraphics[width=0.8\linewidth, trim={5.2cm, 6.3cm, 8.5cm, 1}, clip=true]{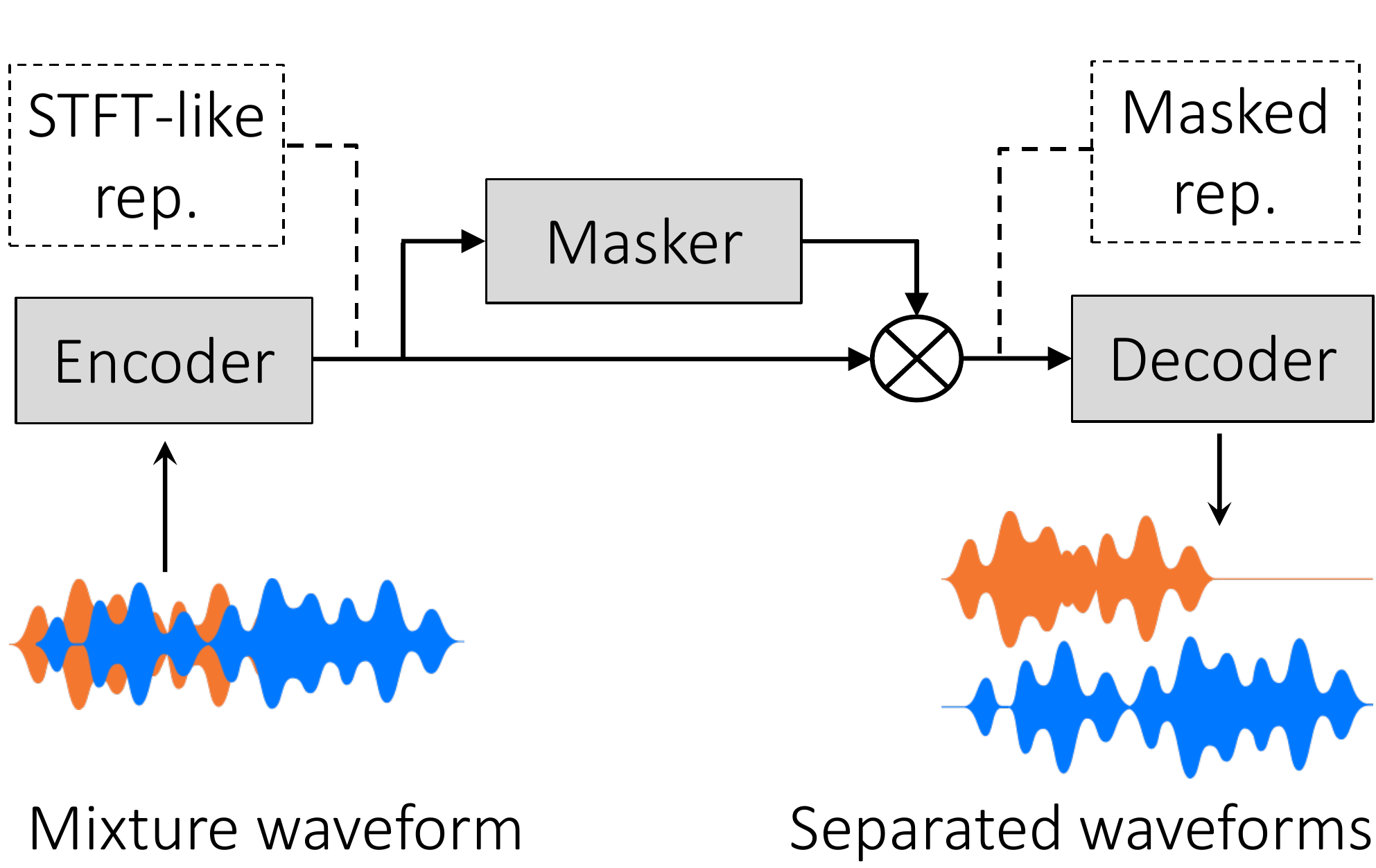}
        \caption{Typical encoder-masker-decoder architecture.}
        \label{fig:encmaskdec}
    \end{figure}

    Most state-of-the-art neural source separation systems follow the encoder-masker-decoder approach depicted
    in Fig.\ \ref{fig:encmaskdec} \cite{LSTMLuo2018, ConvLuo2018, tzinis2019twostep, DPRNNLuo2020}.
    The encoder computes a short-time Fourier transform (STFT)-like representation $\Xmat$ by convolving the time-domain
    signal $x$ with an analysis filterbank. The representation $\Xmat$ is fed to the masker network that estimates a mask for
    each source. 
    The masks are then multiplied entrywise with $\Xmat$ to obtain sources estimates $\{\widehat{\Smat}_j\}_{j=1..J}$ in the
    STFT-like domain. The time-domain source estimates $\{\widehat{s}_j\}_{j=1..J}$ are finally obtained by applying
    transposed convolutions to $\{\widehat{\Smat}_j\}_{j=1..J}$ with a synthesis filterbank. The three networks are 
    jointly trained using a loss function
    computed on the masks or their embeddings \cite{DPCLHershey2016,DPCL+Isik2016,DANetChen2017},
    on the STFT-like domain estimates \cite{PITYu2016, tzinis2019twostep, Demyst2020Heitkaemper},
    or directly on the time-domain estimates \cite{LSTMLuo2018, ConvLuo2018, CompehensiveBahmaninezhad2019,
      Wavesplit2020Zeghidour, DPRNNLuo2020}.

    \section{Functionality}\label{sec:functionality}
    \code{Asteroid} follows the encoder-masker-decoder approach, and provides various choices of filterbanks, masker networks, and loss functions. It also provides training and evaluation tools and recipes for several datasets. We detail each of these below.

    \subsection{Analysis and synthesis filterbanks}\label{subsec:filterbanks}
    As shown in \cite{CompehensiveBahmaninezhad2019, USSKavalerov2019, FilterbankDesign2019Pariente, MultiPhaseDitter2019}, 
    various filterbanks can be used to train end-to-end source separation systems. A natural abstraction is to 
    separate the filterbank object from the encoder and decoder objects. This is what we do in \Asteroid. 
    All filterbanks inherit from the \code{Filterbank} class. Each \code{Filterbank} can be combined 
    with an \code{Encoder} or a \code{Decoder}, which respectively follow the \code{nn.Conv1d} and 
    \code{nn.ConvTranspose1d} interfaces from \code{PyTorch} for consistency and ease of use.
    Notably, the \code{STFTFB} filterbank computes the STFT using simple convolutions, and 
    the default filterbank matrix is orthogonal.

    \code{Asteroid} supports free filters \cite{LSTMLuo2018, ConvLuo2018}, 
    discrete Fourier transform (DFT) filters \cite{USSKavalerov2019, Demyst2020Heitkaemper},
    analytic free filters \cite{FilterbankDesign2019Pariente},
    improved parameterized sinc filters \cite{SincNetRavanelli2018,FilterbankDesign2019Pariente} and 
    the multi-phase Gammatone filterbank \cite{MultiPhaseDitter2019}.
    Automatic pseudo-inverse computation and dynamic filters (computed at runtime) are also supported. 
    Because some of the filterbanks are complex-valued, we provide functions to compute magnitude and phase, and apply 
    magnitude or complex-valued masks. We also provide interfaces to \code{NumPy} \cite{NumPyVanDerWalt2011} and
    \code{torchaudio}\footnote{\href{https://github.com/pytorch/audio}{github.com/pytorch/audio}}.
    Additionally, Griffin-Lim \cite{GriffinLim1984, FastGriffinLim2013} and multi-input spectrogram
    inversion (MISI) \cite{MISI2010} algorithms are provided.

    \subsection{Masker network}
    \code{Asteroid} provides implementations of widely used masker networks:
    TasNet's stacked long short-term memory (LSTM) network \cite{LSTMLuo2018},
    Conv-Tasnet's temporal convolutional network (with or without skip connections) \cite{ConvLuo2018}, and
    the dual-path recurrent neural network (DPRNN) in \cite{DPRNNLuo2020}. Open-Unmix \cite{OpenUnmix} is also supported for music source separation. 

    \subsection{Loss functions --- Permutation invariance}
    \label{subsec:permutation-invariant-losses}
    \code{Asteroid} supports several loss functions: mean squared error, scale-invariant signal-to-distortion 
    ratio (SI-SDR) \cite{ConvLuo2018, SISDRLeroux2019}, scale-dependent 
    SDR \cite{SISDRLeroux2019}, signal-to-noise ratio (SNR), perceptual evaluation of speech quality (PESQ) \cite{PMSQE2018Donas}, and affinity loss for deep clustering \cite{DPCLHershey2016}. 

    Whenever the sources are of the same nature, a permu\-ta\-tion-invariant (PIT) loss shall be used \cite{PITYu2016, uPITKolbaek2017}.
    \code{Asteroid} provides an optimized, versatile implementation of PIT losses.
    Let $\svect = [s_j(t)]_{j=1\dots J}^{t=0\dots T}$ and 
    $\widehat{\svect} = [\widehat{s}_j(t)]_{j=1\dots J}^{t=0\dots T}$ be the matrices of true and estimated source signals, respectively. 
    We denote as $\widehat{\svect}_\sigma = [\widehat{s}_{\sigma (j)}(t)]_{j=1\dots J}^{t=0\dots T}$ a 
    permutation of $\svect$ by $\sigma \in \mathcal{S_J}$, where $\mathcal{S_J}$ 
    is the set of permutations of $[1, ..., J]$.    
    A PIT loss $\mathcal{L}_\text{PIT}$ is defined as 
    \begin{equation}
        \label{PITLoss1}
        \mathcal{L}_\text{PIT}(\theta) = \min_{\sigma \in \mathcal{S_J}} \mathcal{L}(\widehat{\svect}_{\sigma}, \svect), 
    \end{equation}
    where $\mathcal{L}$ is a classical (permutation-dependent) loss function, which depends on the network's 
    parameters $\theta$ through $\widehat{\svect}_{\sigma}$.

    We assume that, for a given permutation hypothesis $\sigma$, the loss $\mathcal{L}(\widehat{\svect}_{\sigma}, \svect)$ can be written as
    \begin{equation}
        \label{PITLossGF}
        \mathcal{L}(\widehat{\svect}_{\sigma}, \svect) = \mathcal{G}\big(\mathcal{F}(\widehat{\svect}_{\sigma(1)}, \svect_1), ..., 
        \mathcal{F}(\widehat{\svect}_{\sigma(J)}, \svect_J)\big)
    \end{equation}
    where $\svect_j=[s_j(0),\dots,s_j(T)]$, $\widehat{\svect}_j=[\widehat{s}_j(0),\dots,\widehat{s}_j(T)]$,
    $\mathcal{F}$ computes the pairwise loss between a single true source and its hypothesized estimate, and 
    $\mathcal{G}$ is the \textit{reduce} function, usually a simple mean operation.
    Denoting by $\Fmat$ the $J\times J$ pairwise loss matrix with entries $\mathcal{F}(\widehat{\svect}_i, \svect_j)$, we can rewrite \eqref{PITLossGF} as
    \begin{equation}
        \label{PITLossGFmat}
        \mathcal{L}_\text{PIT}(\theta) = \min_{\sigma \in \mathcal{S_J}} \mathcal{G}\big(\Fmat_{\sigma(1)1},  ..., 
        \Fmat_{\sigma(J)J}\big)
    \end{equation}    
    and reduce the computational complexity from $J!$ to $J^2$ by pre-computing $\Fmat$'s terms. Taking advantage of this, 
    \code{Asteroid} provides \code{PITLossWrapper}, a simple yet powerful class that can efficiently turn any 
    pairwise loss $\mathcal{F}$ or permutation-dependent loss $\mathcal{L}$ into a PIT loss.

    \subsection{Datasets}
    \code{Asteroid} provides baseline recipes for the following datasets: wsj0-2mix and wsj0-3mix \cite{DPCLHershey2016},
    WHAM \cite{WHAMWichern2019}, WHAMR \cite{WHAMWichern2019}, LibriMix \cite{Librimix2020} FUSS \cite{FUSS2020},
    Microsoft's Deep Noise Suppression challenge dataset (DNS) \cite{DNSChallenge2020}, SMS-WSJ \cite{SMSWSJ2019Drude},
    Kinect-WSJ \cite{KinectWSJ2019}, and MUSDB18 \cite{musdb18}. 
    Their characteristics are summarized and compared in Table \ref{table:datasets}. 
    wsj0-2mix and MUSDB18 are today's reference datasets for speech and music
    separation, respectively. WHAM, WHAMR, LibriMix, SMS-WSJ and Kinect-WSJ are recently released 
    datasets which address some shortcomings of wsj0-2mix. FUSS is the first open-source dataset to 
    tackle the separation of arbitrary sounds.
    Note that wsj0-2mix is a subset of WHAM which is a subset of WHAMR.

    \begin{table*}[t] 
        \centering	
        \setlength\tabcolsep{4pt}
        \small
        \begin{tabular}{l c  c c c c c c c c c}
        \toprule
            & wsj0-mix  & WHAM  & WHAMR & Librimix & DNS  & SMS-WSJ  & Kinect-WSJ & MUSDB18  & FUSS \\
        \midrule
        \midrule
        Source types & speech & speech & speech & speech & speech & speech & speech & music & sounds\\
        \# sources & 2 or 3 & 2 & 2 & 2 or 3 & 1 & 2 & 2 & 4 & 0 to 4 \\
        Noise &  & \Checkmark & \Checkmark & \Checkmark & \Checkmark & * & \Checkmark & & \Checkmark** \\
        Reverb &  &  & \Checkmark &  &  & \Checkmark & \Checkmark & \Checkmark & \Checkmark \\
        \# channels & 1 & 1 & 1 & 1 & 1 & 6 & 4 & 2 & 1 \\
        Sampling rate & 16k & 16k & 16k & 16k & 16k & 16k & 16k & 16k & 16k \\
        Hours & 30 & 30 & 30 & 210 & 100 (+aug.) & 85 & 30 &  10 & 55 (+aug.) \\
        Release year & 2015 & 2019 & 2019 & 2020 & 2020 & 2019 & 2019 & 2017 &  2020 \\
        \midrule
        \bottomrule
        \end{tabular}
            \caption{Datasets currently supported by \code{Asteroid}. * White sensor noise. ** Background environmental scenes.}
    \label{table:datasets}
    \end{table*}
    
    \subsection{Training}
    For training source separation systems, \code{Asteroid} offers a thin wrapper around 
    \code{PyTorch-Lightning} \cite{LightningFalcon2019} that seamlessly enables distributed training,
    experiment logging and more, without sacrificing flexibility.
    Regarding the optimizers, we also rely on native \code{PyTorch} and \code{torch-optimizer}
    \footnote{\href{https://github.com/jettify/pytorch-optimizer}{github.com/jettify/pytorch-optimizer}}.
    \code{PyTorch} provides basic optimizers such as SGD and Adam and \code{torch-optimizer}
    provides state-of-the art optimizers such as RAdam, Ranger or Yogi. 

    \subsection{Evaluation}
    Evaluation is performed using 
    \code{pb\_bss\_eval}\footnote{\href{https://pypi.org/project/pb-bss-eval/}{pypi.org/project/pb\_bss\_eval}}, 
    a sub-toolkit of 
    \code{pb\_bss}\footnote{\href{https://github.com/fgnt/pb_bss}{github.com/fgnt/pb\_bss}} \cite{PB_BSS_Drude} written
    for evaluation. It natively supports most metrics used in source separation: SDR, signal-to-interference ratio (SIR),
    signal-to-artifacts ratio (SAR) \cite{SDRVincent2006}, SI-SDR \cite{SISDRLeroux2019},
    PESQ \cite{PESQRix2001}, and
    short-time objective intelligibility (STOI) \cite{STOITaal2011}.  

    \section{Implementation}
    \label{sec:impl}
\begin{figure}[h]
        \centering
        \includegraphics[width=\linewidth]{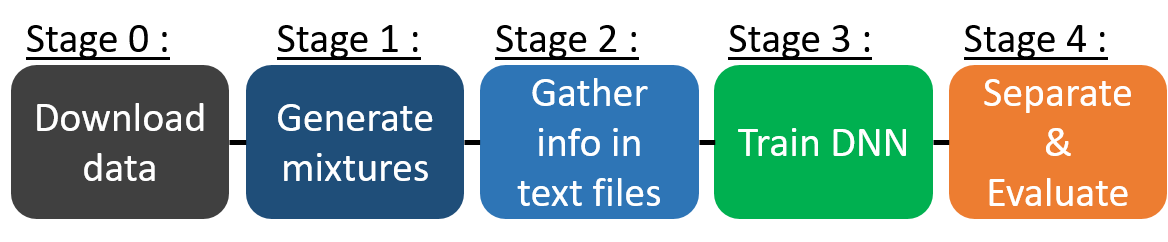}
        \caption{Typical recipe flow in \code{Asteroid}.}
        \label{fig:recipeflow}
    \end{figure}

    \code{Asteroid} follows Kaldi-style recipes  \cite{Povey2011Kaldi}, which involve several stages as depicted 
    in Fig.\ \ref{fig:recipeflow}. These recipes implement the entire pipeline from data download and preparation to
    model training and evaluation. We show the typical organization of a recipe's directory in Fig.\ \ref{fig:dir_structure}.
    The entry point of a recipe is the \code{run.sh} script which will execute the following stages:
    \begin{itemize}
        \item \textbf{Stage 0}: Download data that is needed for the recipe.
        \item \textbf{Stage 1}: Generate mixtures with the official scripts, optionally perform data augmentation.
        \item \textbf{Stage 2}: Gather data information into text files expected by the corresponding \code{DataLoader}.
        \item \textbf{Stage 3}: Train the source separation system.
        \item \textbf{Stage 4}: Separate test mixtures and evaluate. 
    \end{itemize}
    In the first stage, necessary data is downloaded (if available) into a storage directory specified by the user. 
    We use the official scripts provided by the dataset's authors to generate the data, and optionally perform 
    data augmentation. All the information required by the dataset's \code{DataLoader} such as filenames and paths,
    utterance lengths, speaker IDs, etc., is then gathered into text files under \path{data/}. The training stage is
    finally followed by the evaluation stage. Throughout the recipe, log files are saved under \path{logs/} and 
    generated data is saved under \path{exp/}.
    
    \begin{figure}
\begin{lstlisting}[style=tree]
├── data/                 # Output of stage 2
├── exp/                  # Store experiments
├── logs/                 # Store exp logs
├── local/
│ ├── conf.yml            # Training config
│ └── other_scripts.py    # Dataset specific
├── utils/
│ ├── parse_options.sh    # Kaldi bash parser
│ └── other_scripts.sh    # Package-level utils
├── run.sh                # Entry point
├── model.py              # Model definition
├── train.py              # Training scripts
└── eval.py               # Evaluation script
\end{lstlisting}
\caption{Typical directory structure of a recipe.}
\label{fig:dir_structure}
    \end{figure}
    \begin{figure*}[h]
        \centering
        \includegraphics[width=\linewidth, trim={0, 9.5cm, 0, 1}, clip=true]{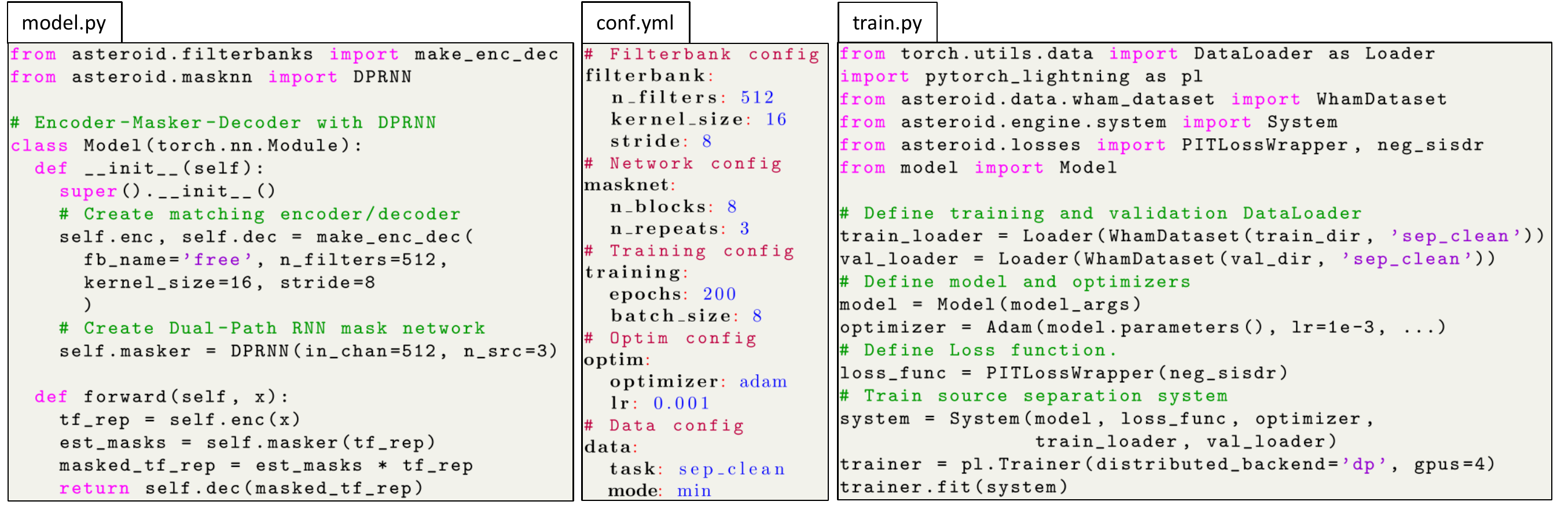}
        \caption{Simplified code example.}
        \label{fig:codexample}
    \end{figure*}
    As can be seen in Fig.\ \ref{fig:codexample},  the model class, which is a direct subclass of \code{PyTorch}'s \code{nn.Module}, is defined in \path{model.py}. It is imported 
    in both training and evaluation scripts.
    Instead of defining constants in \path{model.py} and \path{train.py}, 
    most of them are gathered in a YAML configuration file \path{conf.yml}. 
    An argument parser is created from this configuration file to allow modification of these values from the command 
    line, with \path{run.sh} passing arguments to \path{train.py}.
    The resulting modified configuration is saved in \path{exp/} to enable future reuse.
    Other arguments such as the experiment name, the number of GPUs, etc., are directly passed to \path{run.sh}.

    \section{Example results}\label{sec:experiments}
    To illustrate the potential of \code{Asteroid}, we compare the performance of state-of-the-art methods as reported in the corresponding papers with our implementation. We do so on two common source separation datasets: 
    wsj0-2mix \cite{DPCLHershey2016} and WHAMR \cite{Whamr2019Maciejewski}. 
    wsj0-2mix consists of a 30~h training set, a 10~h validation set, and a 5~h test set of single-channel two-speaker mixtures
    without noise and reverberation. Utterances taken from the Wall Street Journal~(WSJ) dataset are mixed together at random 
    SNRs between $-5$~dB and $5$~dB. Speakers in the test set are different from those in the
    training and validation sets.
    WHAMR \cite{Whamr2019Maciejewski} is a noisy and reverberant extension of wsj0-2mix.
    Experiments are conducted on the 8~kHz \textit{min} version of both datasets. 
    Note that we use wsj0-2mix separation, WHAM's clean separation, and WHAMR's anechoic clean separation tasks 
    interchangeably as the datasets only differ by a global scale.

    Table \ref{table:wham_results} reports SI-SDR improvements (SI-SDR$_\text{i}$) on the test set of wsj0-2mix 
    for several well-known source separation systems. For most architectures, we can see that our implementation outperforms the
    original results. 
    In Table \ref{table:whamr_results}, we reproduce Table 2 from \cite{Whamr2019Maciejewski} which reports the 
    performance of an improved TasNet architecture (more recurrent units, overlap-add for synthesis) on the four main tasks
    of WHAMR: anechoic separation, noisy anechoic separation, 
    reverberant separation, and noisy reverberant separation. On all four tasks, \code{Asteroid}'s recipes achieved 
    better results than originally reported, by up to $2.6$~dB.

    \begin{table}[h!] 
        \centering	
        \setlength\tabcolsep{4pt}
        \small
        \begin{tabular}{l c  c }
        \toprule
            & Reported  & Using \code{Asteroid} \\
        \midrule
        \midrule
        Deep Clustering \cite{DPCLHershey2016} & 10.8 & \\
        TasNet \cite{LSTMLuo2018} & 10.8 & 15.0 \\
        Conv-TasNet \cite{ConvLuo2018} & 15.2 & 16.2 \\
        TwoStep \cite{tzinis2019twostep} & 16.1 & 15.2 \\
        DPRNN ($\text{ks}=16$) \cite{DPRNNLuo2020}& 16.0 & 17.7\\
        DPRNN ($\text{ks}=2$) \cite{DPRNNLuo2020}& 18.8 & 19.3 \\
        Wavesplit \cite{Wavesplit2020Zeghidour} & 20.4 & - \\
        \midrule
        \bottomrule
        \end{tabular}
        \caption{SI-SDR$_\text{i}$ (dB) on the wsj0-2mix test set for several architectures. ks stands for for kernel size, i.e.,
        the length of the encoder and decoder filters.}
    \label{table:wham_results}
    \end{table}
    \begin{table}[H] 
        \centering	
        \setlength\tabcolsep{4pt}
        \small
        \begin{tabular}{c c c  c }
        \toprule
         &   & Reported  & Using \code{Asteroid} \\
         Noise & Reverb   & \cite{Whamr2019Maciejewski} &  \\
        \midrule
        \midrule
         & & 14.2& \bf{16.8}\\
        \Checkmark & & 12.0 & \bf{13.7}\\
        & \Checkmark & 8.9& \bf{10.6}\\
        \Checkmark & \Checkmark & 9.2& \bf{11.0}\\
        \midrule
        \bottomrule
        \end{tabular}
        \caption{SI-SDR$_\text{i}$ (dB) on the four WHAMR tasks using the improved TasNet architecture in 
                \cite{Whamr2019Maciejewski}.}
    \label{table:whamr_results}
    \end{table}
    In both Tables \ref{table:wham_results} and \ref{table:whamr_results}, we can see that our implementations
    outperform the original ones in most cases. Most often, the aforementioned architectures are trained on 4-second segments.
    For the architectures requiring a large amount of memory (e.g., Conv-TasNet and DPRNN), we reduce the length of
    the training segments in order to increase the batch size and stabilize gradients.
    This, as well as using a weight decay of $10^{-5}$ for recurrent architectures increased the final performance of our systems.

    \code{Asteroid} was designed such that writing new code is very simple and results can be quickly obtained.
    For instance, starting from stage 2, writing the TasNet recipe used in Table \ref{table:whamr_results} took 
    less than a day and the results were simply generated with the command in Fig.\ \ref{fig:command_line}, where 
    the GPU ID is specified with the \lstinline{--id} argument.
\begin{figure}[h]
    \centering
\begin{lstlisting}[language=Awk]
n=0
for task in clean noisy reverb reverb_noisy
    do
    ./run.sh --stage 3 --task $task --id $n
    n=$(($n+1))
done
\end{lstlisting}
    \caption{Example command line usage.}
    \label{fig:command_line}
\end{figure}
    \section{Conclusion} \label{sec:conclusion}
    In this paper, we have introduced \code{Asteroid}, a new open-source audio source separation toolkit designed for
    researchers and practitioners. Comparative experiments show that results obtained with \code{Asteroid} are
    competitive on several datasets and for several architectures.
    The toolkit was designed such that it can quickly be extended with new network architectures or new benchmark datasets. In the near future, pre-trained models will be
    made available and we intend to interface with \code{ESPNet} to enable end-to-end multi-speaker speech
    recognition.
    \bibliographystyle{IEEEtran}
    \bibliography{refs}

\end{document}